\documentclass[twocolumn,twoside,slac_two]{revtex4}
\usepackage{graphicx}
\usepackage{fancyhdr}
\begin{document}

\title{Formation and Evolution of Primordial Black Holes After Hybrid Inflation}

%

\author{K.A.~Thompson\footnote{kthom@slac.stanford.edu}
}
\affiliation{Stanford Linear Accelerator Center, Stanford, CA 94309
}

\begin{abstract}
We examine the formation and evolution of primordial black holes (PBH's) after 
hybrid inflation.
 Our goal is to assess the effects of various theoretical
uncertainties on the extrapolation from a given
inflation model to a spectrum of primordial black hole masses.  
The context of our work is an examination of the 
possibility\cite{chenadler03,macgibbon87}
that the dark matter is comprised of Planck-mass black hole remnants (BHR's).
As an example we focus on a particular scenario\cite{chen03a,chen03b} 
in which the black holes form from quantum perturbations that were generated
during hybrid inflation.  
We find the correspondence between hybrid inflation parameters and the range of
initial PBH masses that would allow BHR's to comprise the dark matter,
taking account of the possible early presence of radiation
and its accretion onto the PBH's.
\end{abstract}

\maketitle

\thispagestyle{fancy}


\section{INTRODUCTION}

The main goal of this paper is to assess of
the effects of various theoretical uncertainties on extrapolation from
a given inflation model to a spectrum of primordial black holes.
These uncertainties include initial size of the PBH's, efficiency of 
accretion of ambient radiation onto the PBH's, and possible presence
of radiation due to reheating mechanisms acting prior to PH formation.
As a particular example we examine a
scenario\cite{chen03a,chen03b} in which:  
(1)~significant numbers of PBH's form just after an epoch of hybrid inflation,
(2)~reheating of the Universe occurs primarily through
Hawking evaporation of these PBH's to Planck scale black hole remnants (BHR's), and
(3)~the dark matter consists of these BHR's.
In this BHR dark matter (BHRDM) scenario the problem is further 
simplified because there is a fairly
sharply defined characteristic mass scale at which the PBH's form.
(BHR's as a dark matter candidate were first proposed by MacGibbon\cite{macgibbon87};
other authors have also explored this possibility\cite{barcopelid92,barblaboupol03}).

The endstage of Hawking evaporation is not well understood. However, 
heuristic arguments\cite{adlersant99,adlerchensant01} suggest that 
black holes might not evaporate completely,
but instead may leave behind a stable Planck mass remnant, and that
a black hole's temperature is given by
\begin{eqnarray}
T_{BH} \approx \frac{1}{4\pi M [1+\sqrt{1-\frac{1}{M^2}}]} \;. \label{GUPT}
\end{eqnarray}
rather than the Hawking value
\begin{eqnarray}
T_{BH} \approx \frac{1}{8\pi M} \;. \label{UPT}
\end{eqnarray}
We use Planck units: Planck's constant $\hbar$,
the speed of light $c$, Newton's gravitational constant $G$, 
and the Boltzmann constant $k_b$ are all set to 1.
Thus masses are in units of Planck mass 
$m_P \equiv \sqrt{\hbar c/G} \approx 2.2 \times 10^{-5}$~g 
($\approx 1.2 \times 10^{19}$~GeV),
lengths in units of Planck length 
$l_P \equiv \sqrt{G\hbar/c^3} \approx 1.6 \times 10^{-33}$~cm,
times in units of Planck time 
$t_P \equiv \sqrt{G\hbar/c^5} \approx 5.4 \times 10^{-44}$~sec, etc.

We use a hybrid inflation\cite{linde82,garlinwan96} potential given by:
\begin{equation}
  V(\phi,\psi) = (M^2-{\sqrt{\lambda} \over 2}\psi^2)^2  + {1\over 2}m^2\phi^2 
    + {1\over 2}\gamma \phi^2\psi^2  \quad.
    \label{HybInfPotential}
\end{equation}
The effective mass
of field $\psi$ goes from positive to negative as $\phi$ decreases from large values.
This shift to a negative (``tachyonic'') mass for $\psi$ occurs when 
$\phi$ decreases to the critical value 
$\phi_c = \bigg({2\sqrt{\lambda}M^2 \over \gamma}\bigg)^{1/2}$
There are two inflation regimes in this model: 
(1)~slow rolling of $\phi$ down ``trough'' where $\phi>\phi_c$, with $\psi \approx 0$, 
(2)~Rapid fall of $\psi$ to $\psi_\pm$, beginning when $\phi$ reaches $\phi_c$.
Large perturbations occuring at the ``phase transition between these two regimes
later produce PBH's at a fairly sharply defined mass when they re-enter the horizon.

As discussed in Ref.~\cite{liuthesis04,katliuadchen05}, 
one finds two evolution equations
for the simple ``PBH+radiation epoch'' of a homogeneous, isotropic, flat, 
Friedmann Universe:
\begin{eqnarray}
&& f_{BH}^{\prime} + \frac{f_r^{\prime}}{\bar{a}} = 0,
\label{4EMC2} \nonumber \\
&& f_{BH}^{\prime} =  \frac{\alpha_1 f_{BH}^2 
\frac{f_r}{\bar{a}^4} -
\frac{\alpha_2}{f_{BH}^2}}{\sqrt{\frac{f_{BH}}{\bar{a}} +
\frac{f_r}{\bar{a}^2}}},
\label{4BHRInteraction3} 
\end{eqnarray}
where
\begin{eqnarray}
 \alpha_1 \equiv F \cdot 27\pi \sqrt{\frac{3}{8\pi}}\sqrt{n_i}, \quad 
\alpha_2\equiv \frac{g}{120 \cdot 16^2 \pi}
\sqrt{\frac{3}{8\pi n_i}} \;. \label{alphaDefinition}
\end{eqnarray}
Here $\bar{a} \equiv \frac{a}{a_i}$,
a prime $^{\prime}$ means $d/d\bar{a}$, and we use scaled black hole
and radiation energy densities defined by:
\begin{eqnarray}
 f_{BH} \equiv  \frac{\rho_{BH} \bar{a}^3}{n_i
} = M   \ , \qquad 
 f_r \equiv \frac{\rho_r \bar{a}^4}{n_i} \quad.
\label{4DimensionlessDefinitions}
\end{eqnarray}
where $n_i$ is the initial number density of PBH's, and 
$\rho_{BH} = n_i \bar{a}^{-3} M$ is the energy density in PBH's.
Note that $f_{BH}$ is also equal to the mass $M$ of the PBH's.
The first term in the numerator on the right hand side of
Eq.(\ref{4BHRInteraction3}) represents the accretion of radiation 
(with energy density $\rho_r$)
onto the black holes, and the second term repesents Hawking
evaporation. 
Here we've used the
simpler result in Eq.(\ref{UPT}), and in simulations discussed later will simply
put in a cut-off at the Planck mass at the end of evaporation. 
The factor $g\sim 100$
gives the multiplicity of particles at the high temperatures 
characteristic of black holes near the end of their evaporation.
In obtaining these equations we used the high frequency (``geometrical optics'') 
limit\cite{MTW73} for the absorption cross section 
of the black holes, which is $27 \pi M^2$ for all relativistic particles.
This limit is well-satisfied for the situations considered
in our scenarios.
However, for reasons to be discussed later, we have also introduced
into Eq.~\ref{alphaDefinition}
an ``efficiency factor'' for accretion, $F$.

\section{PBH FORMATION}

\subsection{''Standard picture''}

We begin by reviewing the treatment of PBH formation used in 
Refs.~\cite{chen03a,chen03b}
which is based on the Press-Schechter\cite{pressschect74} type of
formalism put forward by Carr\cite{carr75}.
Assuming spherically symmetric
density perturbations with Gaussian radial profile and rms
amplitude $\delta(M)$, and an equation of state $p=w\rho$ with $0<w<1$, 
Carr argued that the probability $P(m)$ 
of a region of mass $M$ collapsing to form a PBH 
is given by
\begin{equation}
 P(m) \approx \delta(M) \exp\bigg( -\frac{w^2}{2\delta^2(M)} \bigg) \quad. 
\label{eqcarrform}
\end{equation}
Assuming a ``hard'' equation of state (in our case, $p=\rho/3$) 
and a flat Universe the
PBH's are expected to form with approximately the horizon mass.

The dependence of initial conditions
$f_{BH,i}$, $f_{r,i}$ 
(as well as the initial number density of PBH's $n_i$ 
appearing in $\alpha_1$ and $\alpha_2$) 
upon the hybrid inflation potential of Eq.~\ref{HybInfPotential}
can be reduced to two parameters $H_*$ and $s$, where
$H_* \approx \sqrt{8\pi/3}M^2$ is the Hubble parameter during inflation and
$s \equiv -\frac{3}{2}+\sqrt{\frac{9}{4}+\frac{2\sqrt{\lambda}M^2}{H_*^2}}$.
The evolution of $\psi$ during the waterfall regime is given by:
\begin{equation}
  \psi(t) = \psi_{ie} \exp[-s H_* (t-t_{ie})] \quad,
  \label{psiwaterfall}
\end{equation}
where the subscript ``ie'' denotes the end of inflation.

The number of $e$-foldings of inflation between the ``phase transition'' 
near $\phi \sim\phi_c$ and the end of inflation is given by
\begin{equation}
 N_c \approx \bigg[\frac{2}{sH_*} \bigg]^{1/s} \quad.
\end{equation}

The PBH's form when the fractional density perturbations $\delta$
occurring at the ``phase transition'' near $\phi \sim\phi_c$ reenter the horizon.
This density perturbation is closely related to the ``curvature perturbation'' 
$\cal R$ at horizon re-entry\cite{liddlelyth00}:
\begin{equation}
 \delta \equiv \frac{\delta\rho}{\rho} = 
    \frac{2+2w_{reent}}{5+3w_{reent}} {\cal R} \quad.
    \label{deltareent}
\end{equation}
Here $p=w\rho$ is the equation of state, and $w$ at the time of re-entry 
is denoted by $w_{reent}$.
We also have 
\begin{equation}
{\cal R} = \bigg[ \frac{H}{\dot\psi} \delta\psi \bigg]_{t_{reent}} \quad,
\label{curvpert} 
\end{equation}
where $t_{reent}$ denotes that the quantity in brackets is to be evaluated at the time
when the perturbation of interest enters the horizon.
Furthermore, the average spectral perturbation due to quantum fluctuations is
\begin{equation}
|\delta\psi| \sim \frac{H(\psi)}{2\pi} \quad.
\label{qflucts}
\end{equation}


If the Universe is approximately matter-dominated between the end of inflation
and horizon re-entry, then
\begin{equation}
 H_i \approx \frac{2}{3}e^{-3N_c} H_*
  \quad,
\end{equation}
%
%
while if it is more nearly radiation-dominated between the end of inflation
and horizon re-entry, we have
\begin{equation}
 H_i \approx \frac{1}{2}e^{-2N_c} H_*
  \quad.
\end{equation}
%
%

Regardless of where between the two extremes the actual evolution falls,
we assume that the Universe has become radiation dominated by the time 
of re-entry, i.e. $w_{reent}=1/3$. 

The standard assumption is that PBH's form with the horizon mass.
As will be discussed below, this may be only roughly true, 
and changes from this are important 
when considering the effects of accretion onto the PBH's.
So we introduce a factor $x_{BH} (\le 1)$ which gives the ratio 
of the initial PBH Schwarzschild radius 
to the horizon radius $H^{-1}$ at the time of formation:
\begin{equation}
 f_{BH,i} = x_{BH} H_i^{-1}/2
\end{equation}

Finally, the initial proportions of PBH's and radiation (where everything 
except the PBH's falls into the category of ``radiation'')
is obtained as follows.
From Eqs.~\ref{curvpert}, \ref{qflucts}, and \ref{psiwaterfall} we have
${\cal R} \sim 1/s$.
Putting this into Eq.~\ref{deltareent}, gives
\begin{equation}
 \delta_{reent} \equiv \frac{\delta\rho}{\rho} = 
    \frac{2+2w_{reent}}{5+3w_{reent}} \frac{1}{s} = \frac{4}{9s} \quad.
    \label{deltareenttwo}
\end{equation}
Assume the probability $P(M)$ of formation is 
given by Eq.~\ref{eqcarrform}, where $\delta(M)$ 
is given by $\delta_{reent}$ from Eq.~\ref{deltareenttwo}.
(Recall $M$ is
the mass inside the horizon at re-entry, and thus approximately equal to the
mass of the PBH's that form.)
The initial proportions of PBH's and radiation is then just
\begin{equation}
 y \equiv \frac{f_{r,i}}{f_{BH,i}} = \frac{1-P(M) x_{BH}}{P(M) x_{BH}} \quad.
    \label{yprop}
\end{equation}

\subsection{Uncertainties in PBH formation threshold and initial mass}

There remain significant uncertainties in both the threshold overdensity
for PBH formation and in the size
of the PBH's that do form\cite{greenetal04,haracarr04,muscmillrezz04,
nadejinetal78,niemjed9798,novipoln80,shibsasaki99}.  
Early work by Nadejin, Novikov, and Polnarev\cite{nadejinetal78} and
Novikov and Polnarev\cite{novipoln80} gave result that
black holes form with order 10\% the horizon mass 
and also that threshold $\delta$ is higher.
Work by some authors\cite{niemjed9798,shibsasaki99}
indicates critical phenomena in gravitational collapse, 
leading to the possibility of black hole formation with significantly
less than the horizon mass.
Green, et.al.\cite{greenetal04} concluded 
as the results of a study comparing
the Press-Schechter based formalism with a peaks formalism
that the ``standard'' formulation is fairly good if the 
threshold is in the range 0.3 to 0.5.
We do not attempt to settle these uncertainties here,
but note for later use that the initial mass of the PBH's may be 
at least somewhat less than the horizon mass.
\begin{figure}[t]
\centering
\includegraphics[width=70mm]{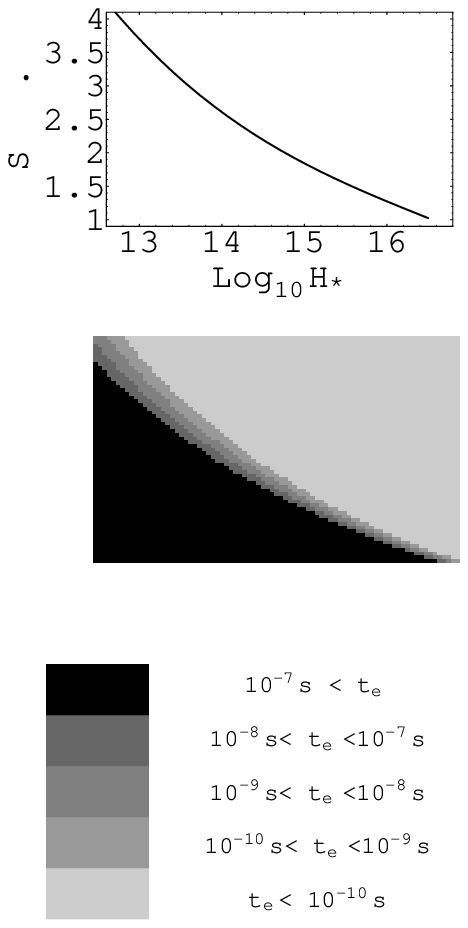}
\caption{Case of no accretion, $w=0$ between the end of inflation and 
the time of PBH formation, and $x_{BH}=1$ 
(baseline case discussed in Refs.~\protect\cite{chen03a,chen03b}).
The top plot shows the line in $H_*,s$ parameter space 
satisfying the constraint that matter-radiation equality occurs
at the observed redshift ($z=3234$).
(Here we display $H_*$ in conventional units of GeV rather than
Planck units.)
The next plot covers the same region in the $H_*,s$ plane and
shows the time $t_e$ at which the evaporation of the
PBH's to BHR's is completed; the gray scale coding in the latter plot is
shown at the bottom of the figure.
} \label{BHfac1_noacc_lpzpzk}
\end{figure}

\section{SIMULATION AND RESULTS}

Given a choice of the hybrid inflation parameters $s,H_*$, we can calculate
initial values $f_{r,i}$, $f_{BH.i}$ and then
solve Eqns.~\ref{4BHRInteraction3} for $f_r(a)$ and $f_{BH}(a)$ up to
the time when the PBH's have evaporated to remnants.
After that, $f_r$ and $f_{BH}$ remain
constant in time (note in particular that $f_{BH}=1$).  
Matter-radiation equality occurs at
\begin{equation}
 \bar a_{eq} = f_{r,e} \quad.
\end{equation}
Here the subscript ``eq'' denotes matter-radiation equality 
and the subscript ``e'' denotes the end of PBH evaporation.
Evolution then continues into the era which is dominated first by dark matter
and eventually, as the present time is approached, by dark energy.
We assume the dark energy behaves like a cosmological constant,
so that $f_{de} \equiv \frac{\rho_{de}}{n_i m_P}$ is constant in time.
(We ignore baryons, in effect just lumping them in with the 
dark matter since they have the same equation of state,
$p \approx 0$.)

The present age of the universe is given by
\begin{equation}
t_0 = t_{eq} + \int_{\bar{a}_{eq}}^{\bar{a}_0} \frac{du}{
\sqrt{\frac{8\pi}{3} n_i \left( \frac{f_{BH,e}}{u} +
\frac{f_{r,e}}{u^2} + f_{de} u^2
\right)}} \;.
\label{t0calc}
\end{equation}
Since the influence of dark energy is negligible until relatively late,
the time of matter-radiation equality is given by
\begin{equation}
t_{eq} =  t_e + \int_{\bar{a}_e}^{\bar{a}_{eq}} \frac{du}{
\sqrt{\frac{8\pi}{3} n_i \left( \frac{f_{BH,e}}{u} +
\frac{f_{r,e}}{u^2} 
\right)}} \;.
\label{teqcalc}
\end{equation}
where $t_e$ is the PBH evaporation time.

Provided that matter-radiation equality occurs in our simulations at the 
observed redshift $z \approx 3234$ 
(and the PBH's do indeed complete their evaporation), the constraint that $t_0$ 
is the observationally inferred value 13.5~Gyr is also satisfied.

\begin{figure}[t]
\centering
\includegraphics[width=70mm]{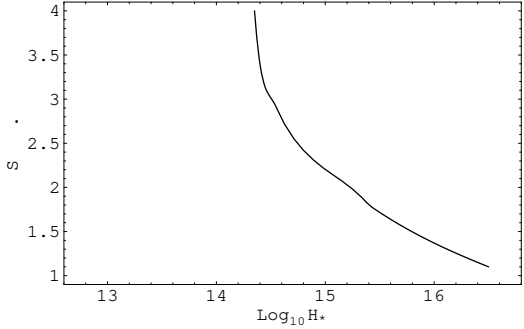}
\caption{Case with accretion included, $w=0$ 
between the end of inflation and 
the time of PBH formation, $x_{BH}=1$, and $F=1$.
The meaning of the plot is as described
for the top plot in Figure~\ref{BHfac1_noacc_lpzpzk}.
{\it Note that the result in this figure cannot be exactly correct, 
since the PBH's initially grow to be larger  
than the horizon -- see Figure~\ref{sgltwofBHandHinv_BHfac1} and discussion in text.} 
} \label{BHfac1_acc_lp}
\end{figure}
\begin{figure}[t]
\centering
\includegraphics[width=70mm]{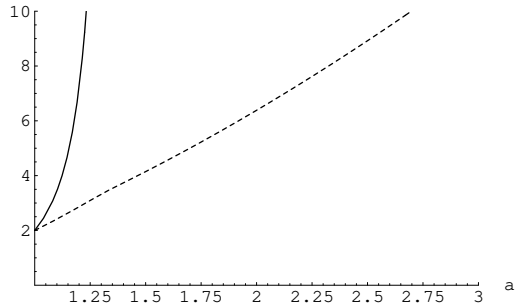}
\caption{For particular point on the line shown in
Figure~\ref{BHfac1_acc_lp} ($s=2.45$, $\log_{10}[H_*/GeV]=14.8$),
we show the PBH radius $2 f_{BH}$ (solid line) 
and the Hubble horizon $H^{-1}$ (dashed line), both normalized by $f_{BH,i}$;
this is as obtained from formulation of accretion
used in Eq.~\ref{4BHRInteraction3} (with $F=1$ in Eq.~\ref{alphaDefinition}).
} \label{sgltwofBHandHinv_BHfac1}
\end{figure}

\subsection{Baseline model}

We begin by showing results for the case discussed in Refs.~\cite{chen03a,chen03b},
for which accretion was not included and 
to which we shall refer as the ``baseline case''.
Thus it was assumed that (for whatever reason) the 
``accretion efficiency factor'' $F$ is essentially zero.
We also assume (1)~black holes form with horizon size ($x_{BH}=1$), 
and (2)~$w \approx 0$ between the end of inflation 
and the time of PBH formation.
The result is shown in Figure~\ref{BHfac1_noacc_lpzpzk}.
The top plot shows the line in $H_*,s$ parameter space 
satisfying the constraint that matter-radiation equality occurs
at the observed redshift ($z=3234$).
The next plot covers the same region in the $H_*,s$ plane and
shows the time $t_e$ at which the evaporation of the
PBH's to BHR's is completed; the gray scale coding in the latter plot is
shown at the bottom of the figure.
A safe criterion would be that they have evaporated by the time of
the electroweak phase transition, expected to be around a TeV, or
$10^{-10}$~sec.
The actual values are
typically somewhat more than this (by one or two orders of magnitude),
but the criterion is somewhat uncertain anyway.

\subsection{Accretion with $\mathbf{F=1}$ and $\mathbf{x_{bh}=1}$}

In Figure~\ref{BHfac1_acc_lp} we include accretion in
accordance with Eq.~\ref{4BHRInteraction3} (with $F=1$ in Eq.~\ref{alphaDefinition}).
There is significant change from the case of no accretion. 
Closer examination shows that in this formulation, where
the black holes are assumed to form with the horizon mass (i.e., $x_{bh}=1$) and 
the accretion efficiency $F=1$, 
the black holes accrete essentially all the available radiation background,
before the scale factor grows appreciably, and 
the early growth rate of the black holes is faster than the
growth of the horizon. 
This is illustrated in
Figure~\ref{sgltwofBHandHinv_BHfac1}
which shows (as a function of scale factor $a$) the PBH radius (solid line) 
and the Hubble horizon $H^{-1}$ (dashed line) 
for the case $s=2.45$, $\log_{10}[H_*/GeV]=14.8$
(which is near the middle of the line in
Figure~\ref{BHfac1_acc_lp}).

As is obvious from causality
(and was long ago pointed out by Zeldovich and Novikov\cite{zeldnovi67})
the mass of the PBH cannot increase faster than the amount of mass within
the sound horizon after formation, which in turn is less than the
optical horizon 
(exactly equal to $H^{-1}$ assuming radiation domination, and in any
case equal up to factor of order 1)

The faster-than-horizon growth
indicates a breakdown of one or more of the assumptions 
in our formulation of accretion, for example,
(1)~that the radiation background remains smooth and uniform, 
and (2)~that the black holes form with the horizon mass.

\subsection{Cases with $\mathbf{x_{bh}=0.4}$}

\begin{figure}[t]
\centering
\includegraphics[width=70mm]{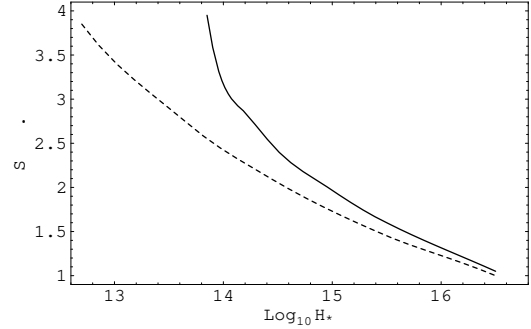}
\caption{Loci in $H_*,s$ parameter space 
satisfying the constraint that matter-radiation equality occurs
at the observed redshift ($z=3234$),
for $x_{bh}=0.4$, with accretion (solid line) and 
without accretion (dashed line);
i.e., same type of plot as in top part of Figure~\ref{BHfac1_noacc_lpzpzk}.
Other parameters are: $w=0$ between the end of 
inflation and the time of PBH formation, and $F=1$ for the case with accretion.
} \label{xbh0.4_accnoacc_lines}
\end{figure}
\begin{figure}[t]
\centering
\includegraphics[width=70mm]{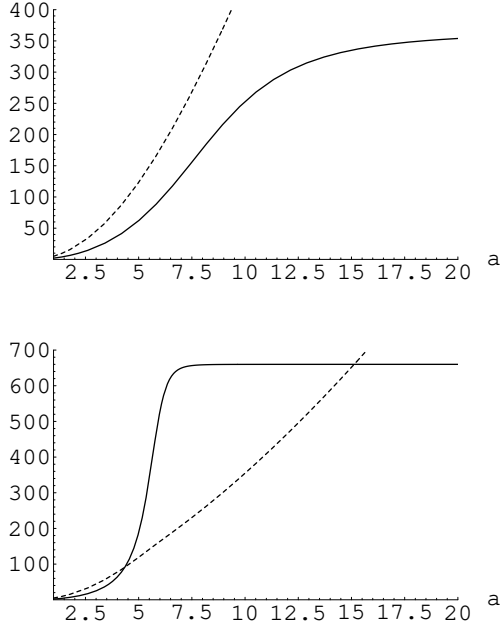}
\caption{The PBH radius $2f_{BH}$ (solid line) 
as a function of scale factor $a$, along
with the Hubble horizon $H^{-1}$ (dashed line), both normalized by $f_{BH,i}$, 
at $s=3.95$, $\log_{10}[H_*/GeV]=13.85$, for $x_{BH}=0.4$ (top plot) and
$x_{BH}=0.41$ (bottom plot).
} \label{sgl_BHfac0.4and0.41_s3.95H13.85}
\end{figure}

As was also pointed out by Zeldovich and Novikov\cite{zeldnovi67},
if PBH's form with significantly less than the horizon mass, accretion is unimportant. 
To get a limit on the possible effect of accretion in the BHRDM scenario,
we reduce $x_{bh}$ just enough to
avoid having PBH's grow to be larger than the Hubble horizon $H^{-1}$.
The required value is $x_{bh}=0.4$, and
Figure~\ref{xbh0.4_accnoacc_lines} 
shows the loci in $H_*,s$ parameter space 
satisfying the constraint that matter-radiation equality occurs
at the observed redshift ($z=3234$),
for this case.  The result with accretion is the solid line, and 
for comparison the dashed line shows the result without accretion.
The effect of accretion is greatest at larger $s$ (upper part of plots)
because the parameter $y \equiv f_{r,i}/f_{BH,i}$ is larger, 
i.e. there is initially more radiation available for accretion. 
In Figure~\ref{sgl_BHfac0.4and0.41_s3.95H13.85}
we illustrate the growth of the PBH's (solid line) and the growth of the 
horizon (dashed line) for a case near the top of the solid line in 
Figure~\ref{xbh0.4_accnoacc_lines}, at $s=3.95$, $\log_{10}[H_*/GeV]=13.85$.
The top plot is for $x_{bh}=0.4$; here the PBH's gain significant mass by accretion,
but never grow faster than the horizon.
The bottom plot is for $x_{bh}=0.41$; here the PBH's do exceed $H^{-1}$ for
a short time.

\subsection{Reduction of accretion efficiency 
}

\begin{figure}[t]
\centering
\includegraphics[width=70mm]{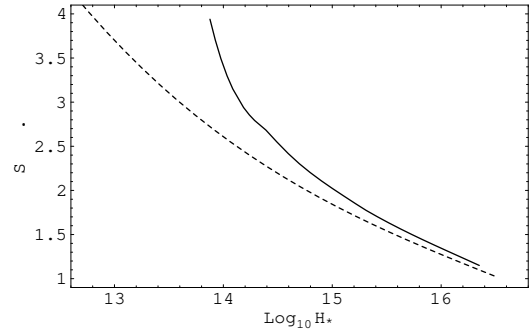}
\caption{Loci in $H_*,s$ parameter space 
satisfying the constraint that matter-radiation equality occurs
at the observed redshift ($z=3234$),
for accretion with $F=0.4$ (solid line) and 
without accretion (dashed line, same as in Figure~\ref{BHfac1_noacc_lpzpzk}).
} \label{Fmod0.4_accnoacc_lines}
\end{figure}
\begin{figure}[t]
\centering
\includegraphics[width=70mm]{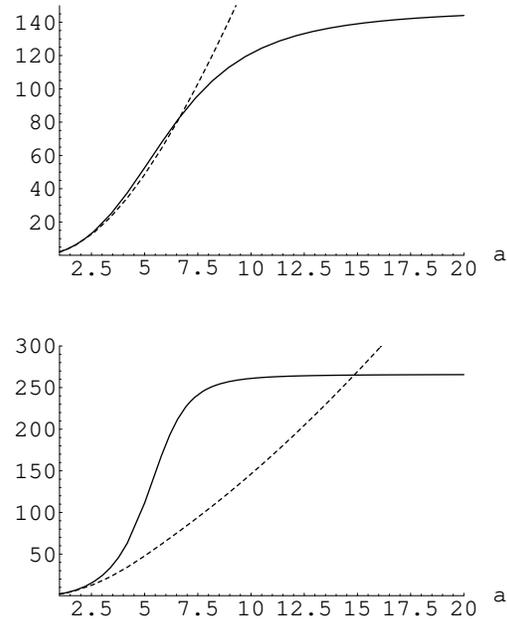}
\caption{The PBH radius $2f_{BH}$ (solid line) 
as a function of scale factor $a$, along
with the Hubble horizon $H^{-1}$ (dashed line), both normalized by $f_{BH,i}$, 
at $s=3.95$, $\log_{10}[H_*/GeV]=13.85$, for $F=0.4$ (top plot) and
$F=0.41$ (bottom plot).
} \label{sgl_Fmod0.4and0.41_s3.95H13.85}
\end{figure}

The solid line in Figure\ref{xbh0.4_accnoacc_lines} represents 
only an upper limit on the PBH growth for this $x=0.4$ case, 
as there is no guarantee that our accretion
formulation is valid even though the growth is not faster than the horizon.
For example,
it could well be the case that the assumption that the radiation background
remains uniform is still violated in reality, i.e. the region around the black
holes could become depleted of radiation, making $F<1$.

It is of course possible to avoid the faster-than-horizon growth 
in the $x_{bh}=1$ case as well, if the accretion efficiency is reduced.
Reduction to $F=0.4$ is sufficient, and results are shown in 
Figures~\ref{Fmod0.4_accnoacc_lines} and \ref{sgl_Fmod0.4and0.41_s3.95H13.85}.
(We would expect $F \rightarrow 1$ eventually, but by that time accretion
becomes neglibible anyway.)

We note Hacyan\cite{hacyan79} found (in a general-relativistic treatment based on
Einstein-Strauss vacuole model) that the 
initial growth of PBH's by radiation accretion could be near the horizon rate 
if the initial PBH size near horizon size, but also
interpreted this result as a generous upper limit on growth rate.

\begin{figure}[t]
\centering
\includegraphics[width=70mm]{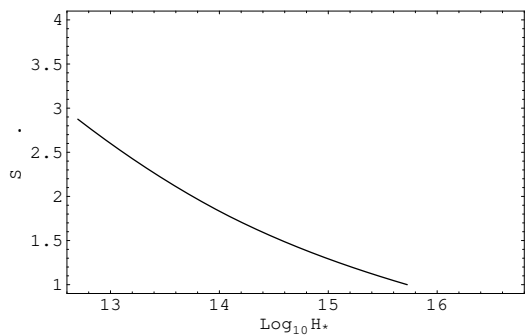}
\caption{Locus in $H_*,s$ parameter space 
satisfying the constraint that matter-radiation equality occurs
at the observed redshift ($z=3234$), for case with no accretion included, $w=1/3$ 
between the end of inflation and 
the time of PBH formation, and $x_{BH}=1$.
} \label{radval_lp}
\end{figure}

\subsection{Possible additional reheating mechanisms}

The baseline case assumed that the Universe is approximately matter 
dominated from the end of inflation until nearly the formation time of the PBH's.
However PBH evaporation is not necessarily the only reheating mechanism.
We need not exclude the possibility of additional reheating mechanism(s)
that could result in approximate radiation domination well before PBH formation.
Thus we compare also a case (Figure~\ref{radval_lp}) where 
$w=1/3$ between the end of inflation and the PBH formation time.
Here we assume other parameters are as in Figure~\ref{BHfac1_noacc_lpzpzk}.
For a given initial $s,H_*$, the $w=1/3$ case gives smaller initial mass of PBH's
than $w=0$, 
since $M \propto 2e^{2N_c}$ instead of $M \propto \frac{3}{2}e^{3N_c}$. 

\section{SUMMARY AND CONCLUSIONS}

We examined effects associated with early presence of radiation
upon extrapolation from hybrid inflation parameters to a spectrum of PBH's 
in the BHRDM model.
It appears that accretion can produce some early rapid growth. 
This may occur even if the requirement that
the PBH's not grow to be larger than the horizon $H^{-1}$ at any time is
enforced by reducing either the fraction $x_{bh}$ of the horizon mass going
into the initial PBH mass or by reducing the accretion efficiency $F$.
However our results should be taken as an upper limit on the effects of
accretion, as there is no guarantee that the efficiency factor is close 
to $F=1$ during the early rapid-accretion phase.
Both these accretion effects and possible radiation production
well before PBH formation affect the correspondence
between the hybrid inflation parameters $s,H_*$ and initial PBH masses,
but do not seriously change the results of the BHRDM scenario.
 
\begin{acknowledgments}
The author thanks Pisin Chen, Ron Adler, Jingsong Liu, Alex Silbergleit, 
and Robert Wagoner for useful discussions. 
Work supported by Department of Energy contract DE-AC03-76SF00515.
\end{acknowledgments}



\begin{thebibliography}{99} 

\bibitem{chenadler03} P.~Chen and R.~Adler, Nucl.Phys.Proc.Suppl. 124, 103 (2003); 
gr-qc/0205106.

\bibitem{macgibbon87} J.H.~MacGibbon, Nature 329, 308 (1987).

\bibitem{chen03a} P.~Chen, astro-ph/0303349. 

\bibitem{chen03b} P.~Chen,  astro-ph/0305025. 

\bibitem{barcopelid92} J.D.~Barrow, E.J.~Copeland, and A.R.~Liddle, PRD 46, 645 (1992).

\bibitem{barblaboupol03} A.~Barrau, D.~Blais, G.~Boudoul, and D.~Polarski, astro-ph/0303330.

\bibitem{adlersant99} R.J.~Adler and D.I.~Santiago, Mod.Phys.Lett. A 14, 
1371 (1999)

\bibitem{adlerchensant01}
R.J.~Adler, P.~Chen and D.~Santiago, Gen.Rel.Grav. 33, 2102 (2001); gr-qc/0106080;

\bibitem{linde82} A.~D.~Linde, Phys.Lett. 108B, 389 (1982).

\bibitem{garlinwan96}
J.~Garcia-Bellido, A.~Linde, and D.Wands, Phys.Rev. D54 (1996) 6040;
astro-ph/9605094.

\bibitem{liuthesis04} J.~Liu, Ph.D. thesis, Stanford Univ., June 2004.

\bibitem{katliuadchen05} K.~Thompson,et.al.,SLAC-PUB-11021,to appear.

\bibitem{MTW73} C.W.~Misner, K.S.~Thorne, and J.A.~Wheeler, {\it Gravitation}, 
Freeman and Co. (1973). 

\bibitem{pressschect74} W.H.~Press and P.~Schechter,
Ap.J.187, 452 (1974).

\bibitem{carr75} B.J.~Carr, Ap.J. 201, 1 (1975).

\bibitem{liddlelyth00} A.R.~Liddle and D.H.~Lyth, {\it Cosmological Inflation
and Large-Scale Structure}, Cambridge University Press, 2000.

\bibitem{greenetal04} A.M.~Green, A.R.~Liddle, K.A.~Malik, and M.~Sasaki; 
astro-ph/0403181.

\bibitem{haracarr04} T.~Harada and B.~Carr, astro-ph/0412134.

\bibitem{muscmillrezz04} I.~Musco, J.C.~Miller, and L.~Rezzolla, gr-qc/0412063.

\bibitem{nadejinetal78} D.K.~Nadejin, I.D.~Novikov, and A.G. Polnarev,
Astron.Zh. 55, 216 (1978). 

\bibitem{niemjed9798} J.Niemeyer and K.Jedamzik, astro-ph/9709072;
J.Niemeyer, astro-ph/9806043.

\bibitem{novipoln80} I.D.~Novikov and A.G.~Polnarev, Astron.Zh. 57, 250 (1980).

\bibitem{shibsasaki99} M.~Shibata and M.~Sasaki, 
PRD 60, 084002 (1999).

\bibitem{zeldnovi67} Ya.B.~Zel'dovich and I.D.~Novikov, 
Soviet Astr-AJ 10, 602 (1967).

\bibitem{hacyan79} S. Hacyan, Ap.J. 229, 42 (1979).

\end{thebibliography}

\end{document}